\documentclass[a4paper,12pt]{article}
\begin{document}
\begin{titlepage}
\begin{flushright}
\begin{tabular}{l}
UTHEP-504\\
KEK-TH-1012\\
hep-th/0503199
\end{tabular}
\end{flushright}

\vspace{5mm}

\begin{center}
{\Large \bf On the Chemical Potential of D-instantons\\ 
in $c=0$ Noncritical String Theory}
\baselineskip=24pt

\vspace{20mm}
\large
Nobuyuki Ishibashi,\footnote{ishibash@het.ph.tsukuba.ac.jp}\\
{\it Institute of Physics, University of Tsukuba,\\
     Tsukuba, Ibaraki 305-8571, Japan}\\
and\\
Atsushi Yamaguchi,\footnote{ayamagu@post.kek.jp}\\
{\it High Energy Accelerator Research Organization (KEK),\\
Tsukuba, Ibaraki 305-0801, Japan }
\vspace{20mm}
\end{center}
\begin{abstract}
We study the chemical potential of D-instantons in $c=0$ noncritical string 
theory. 
In a recent work(hep-th/0405076), it was shown that 
the chemical potential can be calculated using the one matrix model. 
The calculation was done using the method of orthogonal polynomials and the authors obtained a universal value in the double scaling limit. 
We present an alternative method to calculate this value. 
\end{abstract}
\end{titlepage}
\section{Introduction}

Noncritical string theory is a useful toy model in which we can study 
various aspects of critical string theory in a tractable 
way (for reviews, see {\it e.g.} \cite{nc1}\cite{nc2}\cite{nc3}\cite{nc4} and references therein).
In particular, 
we can see that nonperturbative effects of string theory is
of the form $\exp (-S_0/g_s)$\cite{Shenker:1990uf} 
from the analysis of noncritical string
theory, where $g_s$ is the string coupling constant. 
The value of $S_0$ can be deduced from the string equation 
or by using other techniques\cite{David:1990sk}\cite{Eynard:1992sg}\cite{Fukuma:1996hj}. 
This effect can be
considered to be due to D-instantons and $S_0$ 
is the classical action of
the instanton\cite{Polchinski:1994fq}\cite{Zamolodchikov:2001ah}\cite{Alexandrov:2003nn}. 
As in the usual instanton analysis, one should calculate
the chemical potential of the instanton to fix the magnitude of the
nonperturbative effect. 
However, although the form of the nonperturbative
effect can be obtained from the string equation, one cannot fix the
normalization of it from this equation. For a long time, the value of the
chemical potential of the instanton has been unknown. 

Recently, in \cite{hanada},  the authors calculated the chemical potential
for $c=0$ noncritical string theory 
using the matrix model and showed that it is a universal quantity in the
double scaling limit.
This result was generalized to other 
noncritical string theories\cite{Sato:2004tz}\cite{Kawai:2004pj} and was 
further discussed in \cite{deMelloKoch:2004en}. 
In order to study the nonperturbative effects in
the matrix model, it is convenient to consider the effective potential
$V_{eff}(z)$ for the eigenvalue $z$ of the matrix. In \cite{hanada}, the
authors calculated this potential using the method of orthogonal
polynomials. 

In this paper, we would like to present an alternative method to
calculate the chemical potential.
We only need information from the
Schwinger-Dyson equations of the matrix model
and obtain the same value
for the chemical potential as that of \cite{hanada}.
Our method may be
useful to calculate the chemical potential for more generic
type of 
noncritical strings.

The organization of this paper is as follows. In section 2, we
discuss the effective potential $V_{eff}(z)$ of
eigenvalues. We illustrate that if we try to compute $V_{eff}(z)$ 
using the usual genus expansion, the result in
the next to leading order diverges when $z$ is in the region where the
eigenvalues are distributed. In section 3, we explain why the usual
approximation breaks down for such $z$. In section 4, we present a method
to calculate $V_{eff}(z)$ for $z$ in the above mentioned region. The
details of the calculations are given in the appendix. In
section 5, using the results in section 4, we calculate the
value of the chemical potential of the D-instanton and obtain the result
which coincides with the one in \cite{hanada}. Section 6 is devoted to
the conclusions and discussions. 

\section{The effective potential of the matrix eigenvalues}
$c=0$ noncritical string theory can be analyzed by using the one matrix model
\begin{equation}
\int dM e^{-\frac{N}{g^2}{\rm Tr}\, V(M)},
\label{matrix integral}
\end{equation}
where $M$ is an $N\times N$ hermitian matrix and $V(M)$ is a polynomial of 
$M$. Taking the double scaling limit, 
one can calculate various quantities perturbatively with respect to the string 
coupling $g_s$. 
The matrix integral eq.(\ref{matrix integral}) can be expressed by the 
integral 
over the eigenvalues $z_i$ of $M$ as 
\begin{equation}
\int\prod_{i=1}^{N}dz_i
\exp\left(-\frac{N}{g^2}\sum_i V(z_i)+2\mbox{Re}\sum_{i>j}\ln (z_i-z_j)\right).
\end{equation}

In order to study the nonperturbative effects, it is 
convenient to define the effective potential $V_{eff}(z)$ of the matrix 
eigenvalues. 
If one picks up one eigenvalue, say $z_N$, and integrate over 
other eigenvalues putting $z_N=z$, one obtains 
\begin{eqnarray}
& &
\int\prod_{i=1}^{N-1}dz_i
\exp \left(-\frac{N}{g^2}\sum_i^{N-1} V(z_i)+2\mbox{Re}\sum_{i>j}^{N-1}\ln (z_i-z_j)\right.
\nonumber
\\
& &
\hspace{3cm}\left.
-\frac{N}{g^2}V(z)+2\sum_i\ln |z-z_i|\right).
\label{Veff by z}
\end{eqnarray}
This quantity can be regarded as the Boltzmann weight for $z$, and 
integrating it over $z$, we recover the above matrix integral. 
Therefore rewriting eq.(\ref{Veff by z}) in terms of 
an $(N-1)\times (N-1)$ hermitian matrix $M^\prime$, we can define 
$V_{eff}(z)$ so that 
\begin{equation}
e^{-V_{eff}(z)}
=
\frac{
\int dM^\prime 
e^{-\frac{N-1}{g^{\prime 2}}{\rm Tr}\, V(M^\prime )
   -\frac{N-1}{g^{\prime 2}}V(z)+2{\rm Re}{\rm Tr}\ln (z-M^\prime )}
}
{
\int dM^\prime e^{-\frac{N-1}{g^{\prime 2}}{\rm Tr}\, V(M^\prime )}
},
\label{Veff}
\end{equation}
where $g^{\prime 2}=g^2(1-\frac{1}{N})$.

Naively, the right-hand side of the above formula can be calculated as 
\begin{equation}
\exp\left(-\frac{N-1}{g^{\prime 2}}V(z)
      +2\left\langle \mbox{Re}{\rm Tr}\ln (z-M^\prime )\right\rangle 
      +2\left\langle \left(\mbox{Re}{\rm Tr}\ln (z-M^\prime ) \right)^2
      \right\rangle_c +\cdots \right),
\label{Veff by M}
\end{equation}
where 
\begin{equation}
\langle F(M^\prime )\rangle =
\frac{\int dM^\prime e^{-\frac{N-1}{g^{\prime 2}}{\rm Tr}\, V(M^\prime )}F(M^\prime )}
{\int dM^\prime e^{-\frac{N-1}{g^{\prime 2}}{\rm Tr}\, V(M^\prime )}},
\label{langle rangle}
\end{equation}
and the subscript $c$ denotes the connected part. 
Since insertions of ${\rm Tr}\ln (z-M^\prime )$ can be regarded as generating 
boundaries on the worldsheet, eq.(\ref{Veff by M}) can be considered as 
a genus expansion of the free energy in open string theory. 
Therefore this expansion is natural, considering the relation 
between the 
D-branes and the eigenvalues 
in the matrix model\cite{FZZT}\cite{McGreevy:2003kb}. 
However, as we will see, this expansion is not always valid. 

Eq.(\ref{Veff by M}) seems to be valid in the leading order of 
$N\rightarrow\infty$ limit. 
Indeed, using this formula, the leading order contribution to 
$V_{eff}(z)$ is given as 
\begin{equation}
V_{eff}(z)
=
-(N-1)\left(\frac{2}{N-1}\left\langle \mbox{Re}{\rm Tr} \ln (z-M^\prime )\right\rangle 
  -\frac{1}{g^{\prime 2}}V(z)\right).
\end{equation}
An explicit form of this potential can be obtained once we know 
the resolvent 
$w(z)\equiv\frac{1}{N-1}\langle {\rm Tr}\frac{1}{z -M^\prime}\rangle $ 
in the large $N$ limit. 
$w(z)$ can be obtained by solving the Schwinger-Dyson equation as
\begin{equation}
w(z)
=
\frac{1}{2g^{\prime 2}}
\left(V^\prime (z)-M(z)\sqrt{(z-\alpha )(z-\beta )}\right),
\label{wz}
\end{equation}
where $M(z)$ is a polynomial. 
$M(z)$ and $\alpha ,~\beta ~(\beta >\alpha )$ are 
fixed by the condition $w(z)\sim {1}/{z}$ when $z\rightarrow\infty$. 
Thus we get $V_{eff}(z)$ as 
\begin{equation}
(N-1)
\int^zdz^\prime 
\frac{1}{g^{\prime 2}}M(z)\mbox{Re}\sqrt{(z-\alpha )(z-\beta )}.
\end{equation}
This potential is flat for the cut $\beta \geq z\geq\alpha $ of $w(z)$. 
This is consistent with the fact that the eigenvalues are distributed 
there. If we define $\rho (z)$ to be the distribution function of 
the eigenvalues of $M^\prime $, $w(z)$ can be expressed by $\rho(z)$ as 
\begin{equation}
w(z)=\int dz^\prime \frac{\rho (z^\prime )}{z-z^\prime}.
\label{wzrho}
\end{equation}
The inverse relation is $\rho (z)=-\frac{1}{\pi}\mbox{Im}\, w(z)$. Therefore we can see that the eigenvalues 
are distributed in  the cut $\beta \geq z\geq \alpha $ of $w(z)$. 

$V_{eff}(z)$ has a local maximum at $z=\gamma $ near $z=\beta$ and the 
configuration where one of the eigenvalues is at this point can be 
regarded as the D-instanton configuration. 
\footnote{In this paper, we assume that $V(z)$ is not an even polynomial, 
for simplicity.} 
D-instantons contribute to the partition function and one can evaluate such 
nonperturbative corrections to the free energy as\cite{hanada}
\begin{equation}
\delta F
=
\frac{N\int_{\beta}^\infty dz\exp \left(-V_{eff}(z)\right)}
{\int^{\beta}_\alpha dz\exp \left(-V_{eff}(z)\right)}.
\label{deltaF}
\end{equation}
$N$ in the numerator is obtained by considering which of the $N$ eigenvalues 
are at $z=\gamma$. 
In the double scaling limit, $\delta F$ becomes of the form 
$\exp (-{S_0}/{g_s}+\ln\mu_{inst.}+{\cal O}(g_s))$. 
$S_0$, $\mu_{inst.}$ can be considered as the classical action and 
the chemical potential of the D-instanton respectively. 
One can evaluate $S_0$ from $V_{eff}(z)$ in the large $N$ limit given above, 
and get the value consistent with the one from the string equation. 

In order to get the chemical potential of the D-instanton, we should consider 
the next to leading order contribution in the large $N$ limit. If we trust 
eq.(\ref{Veff by M}) what we need is the two loop correlator in the matrix 
model. It is known that\cite{Ambjorn:1990ji} 
\begin{equation}
\left\langle {\rm Tr}\frac{1}{z-M^\prime}{\rm Tr}\frac{1}{z^\prime -M^\prime}\right\rangle_c  
=
\frac{1}{4(z-z^\prime )^2}
\left[
\frac{2zz^\prime -(\alpha +\beta )(z+z^\prime )+2\alpha\beta }
{\sqrt{(z-\alpha )(z-\beta )}\sqrt{(z^\prime -\alpha )(z^\prime -\beta )}}
-2
\right].
\label{two loop}
\end{equation}
This quantity is universal in the sense that it depends on $V(z)$ only through 
the values of $\alpha ,~\beta$. Using this formula, we obtain 
\begin{equation}
\left\langle {\rm Tr}\ln (z-M^\prime ){\rm Tr}
\ln (z^\prime -M^\prime )\right\rangle_c  
=
-\ln 
\left(1-e^{-\theta -\theta^\prime}\right),
\end{equation}
where the variable $\theta$ is defined to be 
\begin{equation}
e^{-\theta}=\frac{2}{\beta -\alpha }
\left(z-\frac{\alpha +\beta }{2}-\sqrt{(z-\alpha )(z-\beta )}\right).
\end{equation}

Therefore the next to leading order contribution to $V_{eff}(z)$ is given as 
\begin{equation}
\left.
2\mbox{Re}_\theta \mbox{Re}_{\theta^\prime}
\ln (1-e^{-\theta -\theta^\prime})\right|_{z=z^\prime}.
\end{equation}
The behavior of this quantity depends crucially on whether $z$ is 
inside the cut ($\beta \geq z\geq \alpha $) or outside the cut 
($z>\beta$ or $\alpha >z$). If it is inside the cut, 
$\theta$ is imaginary and we get 
\begin{eqnarray}
\left.2\mbox{Re}_\theta \mbox{Re}_{\theta^\prime}
\ln (1-e^{-\theta -\theta^\prime})\right|_{z=z^\prime}
&=&
\frac{1}{2}
\ln \left[(1-e^{-\theta -\theta^\prime})(1-e^{-\theta +\theta^\prime})\right.
\nonumber
\\
& &
\hspace{15mm}
\left. \left.\times
     (1-e^{\theta -\theta^\prime})
(1-e^{\theta +\theta^\prime})\right]
     \right|_{\theta =\theta^\prime}
\nonumber
\\
&=&
\infty .
\end{eqnarray}
Therefore the next to leading order contribution to $V_{eff}(z)$ in 
eq.(\ref{Veff by M}) diverges. 
In order to avoid this divergence, the authors of \cite{hanada} used the
method of orthogonal polynomials instead of the naive formula
eq.(\ref{Veff by M}). 
Actually such a divergence is necessary for eq.(\ref{deltaF}) to have 
a finite value in the double scaling limit. Since $N\rightarrow\infty$, 
we need an factor proportional to $N$ in the denominator. 
Any way, we cannot use eq.(\ref{Veff by M}) to calculate such a factor. 
For $z$ outside the cut, 
\begin{equation}
 \left.2\mbox{Re}_\theta \mbox{Re}_{\theta^\prime}
\ln (1-e^{-\theta -\theta^\prime})\right|_{z=z^\prime}
=
2
\ln (1-e^{-2\theta}),
\label{inside}
\end{equation}
and we do not have such a problem.

The existence of the divergence indicates that the expansion 
eq.(\ref{Veff by M}) is not a valid approximation for 
$\beta \geq z\geq \alpha$, 
while it has no problems for $z>\beta ,~\alpha >z$. 
We would like to discuss why the expansion eq.(\ref{Veff by M}) is not
valid for $\beta \geq z\geq\alpha$, in the next section. 

\section{Eq.(\ref{Veff by M}) for $\beta \geq z\geq\alpha$}

In order to understand the reason why eq.(\ref{Veff by M}) is not valid
for $\beta \geq z\geq\alpha$, let us study the matrix 
integral 
\begin{equation}
\int dM^\prime 
e^{-\frac{N-1}{g^{\prime 2}}{\rm Tr}\,V(M^\prime )
   -\frac{N-1}{g^{\prime 2}}V(z)+2{\rm Re}{\rm Tr}\ln (z-M^\prime )},
\label{matrix model}
\end{equation}
without expanding the integrand as eq.(\ref{Veff by M}). 
This can be considered as a matrix model with a matrix potential 
$\frac{N-1}{g^{\prime 2}}{\rm Tr}V(M^\prime )-2{\rm Re}{\rm Tr}\ln (z-M^\prime )$. 

To study such a model, we consider the resolvent 
\begin{equation}
\tilde{w}(z^\prime )
\equiv
\frac{1}{N-1}\left\langle {\rm Tr}\frac{1}{z^\prime -M^\prime}\right\rangle^\prime ,
\end{equation}
where $\langle \cdot\rangle^\prime$ denotes the expectation value in this 
matrix model. 
Notice that $\tilde{w}(z^\prime )$ implicitly depends on $z$. 
The Schwinger-Dyson equation satisfied by the resolvent can be constructed 
in the usual way, and we obtain in the large $N$ limit, 
\begin{eqnarray}
&&\hspace{-30pt}\left(\tilde{w}(z^\prime )\right)^2
-\frac{1}{g^{\prime 2}}V^\prime (z^\prime )\tilde{w}(z^\prime )\nonumber\\
&&\hspace{-10pt}+\frac{1}{N-1}
\left(\frac{\tilde{w}(z^\prime )-\tilde{w}(z+i\epsilon )}
      {z^\prime -z-i\epsilon}
+
 \frac{\tilde{w}(z^\prime )-\tilde{w}(z-i\epsilon )}
      {z^\prime -z+i\epsilon}\right)
=\frac{1}{4g^{\prime 4}}f(z^\prime ),
\label{SD equation}
\end{eqnarray}
where $f(z^\prime )$ is a polynomial of $z^\prime $. 

Now let us consider how the eigenvalues of $M^\prime$ are distributed in the 
matrix model eq.(\ref{matrix model}). Since there exists 
$-2{\rm Re}{\rm Tr}\ln (z-M^\prime )$ in the potential, eigenvalues cannot be 
distributed near $z$. Therefore $\tilde{w}(z)$ cannot have any imaginary part 
and $\tilde{w}(z+i\epsilon )=\tilde{w}(z-i\epsilon )=\tilde{w}(z)$. 
Thus, for $z^\prime $ not on the real axis, we get 
\begin{equation}
(\tilde{w}(z^\prime ))^2
-\frac{1}{g^{\prime 2}}V^\prime (z^\prime )\tilde{w}(z^\prime )
+\frac{2}{N-1}\frac{\tilde{w}(z^\prime )}{z^\prime -z}
=
\frac{2}{N-1}\frac{\tilde{w}(z)}{z^\prime -z}
+\frac{1}{4g^{\prime 4}}f(z^\prime ), 
\end{equation}
which can be solved as
\begin{eqnarray}
\tilde{w}(z^\prime )
&=&
\frac{1}{2 g^{\prime 2}}V^\prime (z^\prime )
-\frac{1}{N-1}\frac{1}{z^\prime -z}
\nonumber
\\
& &
\hspace{5mm}
+
\sqrt{\frac{1}{4g^{\prime 4}}\left(
\left(V^\prime (z^\prime )\right)^2
+f(z^\prime )\right)
      +\frac{2}{N-1}\frac{\tilde{w}(z)}{z^\prime -z}
      +\left(\frac{1}{N-1}\frac{1}{z^\prime -z}\right)^2}.
\nonumber
\\
& &
\label{wtilde}
\end{eqnarray}
In this expression we can see that for $z^\prime$ very close to $z$,
$(\frac{1}{N-1}\frac{1}{z^\prime -z})^2$ becomes very big and
$\tilde{w}(z^\prime )$ has no imaginary part. 
In order to obtain $\tilde{w}(z^\prime )$ we should fix $f(z^\prime )$ and 
$\tilde{w}(z)$. Such quantities are usually determined by the conditions 
that $\tilde{w}(z)$ has singularities expected from the eigenvalue 
distributions. 

The matrix model we are considering are made by picking up one eigenvalue 
of $M$ and putting it at $z$. Therefore we expect that the distribution of 
the eigenvalues should be not so different from the original one, i.e. 
$\rho (z^\prime )$. When $z>\beta ,~\alpha >z$ this will be the case. 
Since $z$ is outside the cut, the potential 
$-2{\rm Re}{\rm Tr}\ln (z-M^\prime )$ distorts the original distribution but the 
effect is of ${\cal O}(\frac{1}{N})$. 
Thus $f(z^\prime )$ and 
$\tilde{w}(z)$ can be fixed by the condition that $\tilde{w}(z^\prime )$ 
has one cut. 
On the other hand, for $z$ with $\beta \geq z\geq\alpha$, the situation is 
quite different. Since there cannot be any eigenvalues near $z$, the original 
distribution is divided into two. Hence we should solve eq.(\ref{SD equation}) 
with the condition that $\tilde{w}(z^\prime )$ has two cuts. 

Actually we can expand the resolvent $\tilde{w}(z^\prime )$ as in
eq.(\ref{Veff by M}) as 
\begin{equation}
\tilde{w}(z^\prime )
=
\sum_n\frac{1}{n!}
\frac{1}{N-1}
\left\langle {\rm Tr}\frac{1}{z^\prime -M^\prime}
\left(2{\rm Re}{\rm Tr}\ln (z-M^\prime )\right)^n
\right\rangle_c ,
\end{equation}
where $\langle \cdot \rangle$ is defined in eq.(\ref{langle rangle}). 
If we expand $\tilde{w}(z^\prime )$ as 
\begin{equation}
\tilde{w}(z^\prime )
=
\tilde{w}_0(z^\prime )
+\frac{1}{N-1}\tilde{w}_1(z^\prime )
+\frac{1}{(N-1)^2}\tilde{w}_2(z^\prime )
+\cdots ,
\label{wn}
\end{equation}
we can expand eq.(\ref{SD equation}) in terms of $\frac{1}{N-1}$ 
and we can see that $\tilde{w}_n(z^\prime )$ and 
$(N-1)^n\langle {\rm Tr}\frac{1}{z^\prime -M^\prime}
(2{\rm Re}{\rm Tr}\ln (z-M^\prime ))^n\rangle_c$ satisfy the same equations. 
When $z>\beta ,~\alpha >z$, we
can conclude that 
\begin{equation}
\tilde{w}_n(z^\prime )
=
(N-1)^n\left\langle {\rm Tr}\frac{1}{z^\prime -M^\prime}
\left(2{\rm Re}{\rm Tr}\ln (z-M^\prime )\right)^n\right\rangle_c,
\end{equation}
because the condition that $\tilde{w}(z^\prime )$ has one cut coincides
with the ones for $(N-1)^n\langle {\rm Tr}\frac{1}{z^\prime -M^\prime}
(2{\rm Re}{\rm Tr}\ln (z-M^\prime ))^n\rangle_c$. 
Therefore we can treat the term 
$2{\rm Re}{\rm Tr}\ln (z-M^\prime )$ perturbatively. Since one can calculate 
the partition function from $\tilde{w}(z^\prime )$, we conclude that we 
can use eq.(\ref{Veff by M}) in this case. However, for $z$ with $\beta
>z>\alpha$, this is not the case because the conditions for the cut is
different. It implies that we cannot treat 
$2{\rm Re}{\rm Tr}\ln (z-M^\prime )$ 
perturbatively and therefore we cannot trust eq.(\ref{Veff by M}) 
when $\beta \geq z\geq \alpha$.

To sum up, when $z>\beta ,~\alpha >z$ we can use eq.(\ref{Veff by M}). 
For $z$ with $\beta \geq z\geq\alpha$ we should not use 
eq.(\ref{Veff by M}) which treats $2{\rm Re}{\rm Tr}\ln (z-M^\prime )$ 
perturbatively. 
In this case, for $z^\prime$ close to $z$ the effect of the potential 
$2{\rm Re}{\rm Tr}\ln (z-M^\prime )$ is not perturbative. 
In principle, we can calculate $V_{eff}(z)$ using the 
matrix model eq.(\ref{matrix model}) when $\beta \geq z\geq\alpha$. 
\footnote{
We can fix the value of $\tilde{w}(z)$ from the condition that 
the number of the eigenvalues larger than $z$ is the same as that of the 
original matrix model.} 
However, since the next to leading order calculations of the matrix model 
eq.(\ref{matrix model}) are quite complicated, we resort to a different method.

\section{$V_{eff}(z)$ for $\beta \geq z\geq\alpha$}

Since $V_{eff}(z)$ cannot be calculated using eq.(\ref{Veff by M}) for 
$\beta \geq z\geq \alpha$, 
we may not be able to trust the conclusions drawn from 
the leading order analysis in section 2. 
However, the $z$ dependence of $V_{eff}(z)$ can be found from the following 
arguments, and the properties of $V_{eff}(z)$ given in section 2 turns out 
to be roughly true.  
Let us consider 
\begin{equation}
\int dM e^{-\frac{N}{g^2}V(M)}\frac{1}{N}{\rm Tr}\frac{1}{z-M}.
\label{resolvent}
\end{equation}
Eq.(\ref{resolvent}) can be rewritten into an integral over 
matrix eigenvalues as
\begin{eqnarray}
& &
\int\prod_{i=1}^{N}dz_i
e^{-\frac{N}{g^2}\sum V(z_i)+2{\rm Re}\sum_{i>j}\ln (z_i-z_j)}
\frac{1}{N}\sum_i\frac{1}{z-z_i}
\nonumber
\\
& &
\hspace{5mm}
=
\int\prod_{i=1}^{N}dz_i
e^{-\frac{N}{g^2}\sum V(z_i)+2{\rm Re}\sum_{i>j}\ln (z_i-z_j)}
\frac{1}{z-z_N}
\nonumber
\\
& &
\hspace{5mm}
=
C
\int dz^\prime e^{-V_{eff}(z^\prime )}\frac{1}{z-z^\prime},
\end{eqnarray}
where $C$ is some constant. 
Since the distribution function of the eigenvalues of $M$ can be obtained 
by taking the imaginary part of eq.(\ref{resolvent}) in the large $N$ limit, 
we can see that $\exp (-V_{eff}(z))$ is proportional to 
this distribution function in the large $N$ limit. 
In the large $N$ limit, the distribution function for $M$ coincides with 
that for $M^\prime$, i.e. $\rho (z)$. 
Indeed in \cite{hanada} this fact was shown as a result of the
calculation of $\exp (-V_{eff}(z))$. 
Therefore we can conclude that  
$\exp (-V_{eff}(z))$ is very small outside the cut. 
Since we can use eq.(\ref{Veff by M}) outside the cut, $V_{eff}(z)$ 
possesses a local maximum near the cut and we can identify the D-instanton 
configuration. 

In order to evaluate the chemical potential of the instanton, 
we need to calculate 
$\int_\alpha^\beta 
dze^{-V_{eff}(z)}\sim \int dze^{-V_{eff}(z)}$. 
Let us obtain this quantity through rather brute force calculations. 
From the definition eq.(\ref{Veff}) we get 
\begin{equation}
\int dze^{-V_{eff}(z)}
=
\frac{\int dz
\int dM^\prime 
e^{-\frac{N-1}{g^{\prime 2}}{\rm Tr}\, V(M^\prime )
   -\frac{N-1}{g^{\prime 2}}V(z)+2{\rm Re}{\rm Tr}\ln (z-M^\prime )}
}
{
\int dM^\prime e^{-\frac{N-1}{g^{\prime 2}}{\rm Tr}\, V(M^\prime )}
}.
\label{N/N-1}
\end{equation}
Since the numerator is proportional to the partition function 
$\int dM e^{-\frac{N}{g^2}{\rm Tr}V(M)}$, what we should calculate is 
essentially the ratio between the matrix integrals  over
$M$ and $M^\prime$. 
These matrix integrals can be evaluated using $\rho (z)$ and $w(z)$ in the 
large $N$ limit. After some calculations, which are presented in the appendix, 
we obtain 
\begin{equation}
\int dze^{-V_{eff}}
\cong 
\frac{\pi (N-1)}{2}(\beta -\alpha )e^{2(N-1)R},
\label{dzVeff}
\end{equation}
where
\begin{equation}
R
\equiv
\lim_{\Lambda\rightarrow\infty}
\left[\int_\Lambda^\beta dz^\prime w(z^\prime )
+\ln \Lambda \right]-\frac{1}{2g^{\prime 2}}V(\beta ).
\label{R}
\end{equation}
From eq.(\ref{dzVeff}), we can see that for $\beta \geq z\geq\alpha $, 
\begin{equation}
e^{-V_{eff}(z)}
\cong
\frac{\pi (N-1)}{2}(\beta -\alpha )e^{2(N-1)R}\rho (z).
\end{equation}

\section{The chemical potential of the instanton}
Now that we know $V_{eff}(z)$ when $\beta \geq z\geq\alpha $, we can calculate the 
chemical potential of the instanton. For $z$ larger than $\beta$ we can 
use the formula eq.(\ref{Veff by M}) and we obtain
\begin{eqnarray}
e^{-V_{eff}(z)}
&\cong& 
\frac{1}{(1-e^{-2\theta})^2}
\exp 
\left[2(N-1)R-
\frac{N-1}{g^{\prime 2}}\int^z_\beta dz^\prime M(z^\prime )
\sqrt{(z^\prime -\alpha )(z^\prime -\beta )}\right],\nonumber\\
& &
\end{eqnarray}
in the large $N$ limit. 

The integral $\int_{\beta}^\infty dze^{-V_{eff}(z)}$ can be evaluated using 
the saddle point approximation. 
The saddle point we are interested in is 
$z=\gamma$ near the cut, at which $V_{eff}(z)$ has a local maximum. 
In the large $N$ limit $V_{eff}(z)\sim 
\frac{N-1}{g^{\prime 2}}\int^z_\beta dz^\prime M(z^\prime )
\sqrt{(z^\prime -\alpha )(z^\prime -\beta )}$ and the polynomial 
$M(z^\prime )$ is factorized as 
\begin{equation}
-\frac{1}{g^{\prime 2}}M(z^\prime )=G(z^\prime )(z^\prime -\gamma ). 
\end{equation}
Then 
\begin{eqnarray}
\int_{\beta}^\infty dze^{-V_{eff}(z)}
&\cong &
i\sqrt{
\frac{2\pi}
{(N-1)G(\gamma )\left((\gamma -\alpha )(\gamma -\beta )
\right)^{\frac{5}{2}}}
}
\nonumber
\\
& &
\hspace{2mm}
\times
\frac{(\beta -\alpha )^4}
{64\left(\gamma -\frac{\alpha +\beta }{2}-
\sqrt{(\gamma -\alpha )(\gamma -\beta )}\right)^2}
\nonumber
\\
& &
\hspace{2mm}
\times
\exp 
\left[2(N-1)R-
\frac{N-1}{g^{\prime 2}}\int^\gamma_\beta dz^\prime M(z^\prime )
\sqrt{(z^\prime -\alpha )(z^\prime -\beta )}\right].
\nonumber
\\
& &
\end{eqnarray}

Thus the instanton contribution to the partition function is given as 
\begin{eqnarray}
\delta F
&=&
\frac{N\int_\beta^\infty dze^{-V_{eff}(z)}}
{\int dze^{-V_{eff}(z)}}
\nonumber
\\
&\cong&
\frac{i(\beta -\alpha )^3}
{16\sqrt{2\pi}
\left
(\gamma -\frac{\alpha +\beta }{2}-\sqrt{(\gamma -\alpha )(\gamma -\beta
)}
\right)^2
\sqrt{NG(\gamma )((\gamma -\alpha )(\gamma -\beta ))^{\frac{5}{2}}}
}
\nonumber
\\
& &
\hspace{5mm}
\times
\exp 
\left[-
\frac{N-1}{g^{\prime 2}}\int^\gamma_\beta dz^\prime M(z^\prime )
\sqrt{(z^\prime -\alpha )(z^\prime -\beta )}\right].
\end{eqnarray}

In the continuum limit, we take the limit in which $g^{\prime 2}$ goes 
to the critical value $g_c^{\prime 2}$, which satisfies  
$\beta (g_c^{\prime 2})=\gamma (g_c^{\prime 2})=z_c$. 
More explicitly, we take the limit $a\rightarrow 0$ so that 
\begin{equation}
g_c^{\prime 2}-g^{\prime 2}\propto \mu a^2,
\end{equation}
where $\mu$ is the cosmological constant. Then with an appropriate 
normalization of $\mu$, 
\begin{eqnarray}
\beta 
&\cong&
z_c-\sqrt{\mu}a,
\nonumber
\\
\gamma
&\cong&
z_c+\frac{1}{2}\sqrt{\mu}a.
\end{eqnarray}
In this limit, the classical action part of the instanton becomes 
\begin{equation}
\frac{N-1}{g^{\prime 2}}\int^\gamma_\beta dz^\prime M(z^\prime )
\sqrt{(z^\prime -\alpha )(z^\prime -\beta )}
\cong
\frac{8\sqrt{3}}{5}\mu^{\frac{5}{4}}
\left(3\cdot 2^{-\frac{7}{2}}NG(z_c)\sqrt{z_c-\alpha }a^{\frac{5}{2}}\right).
\end{equation}
Therefore if we take the double scaling limit $N\rightarrow\infty$, 
$a\rightarrow 0$ with 
\begin{equation}
 3\cdot 2^{-\frac{7}{2}} N G(z_c)
\sqrt{z_c-\alpha } a^{\frac{5}{2}} = \frac{1}{g_s}
\end{equation}
fixed, 
we get 
\begin{equation}
\frac{8\sqrt{3}\mu^{\frac{5}{4}}}{5g_s},
\end{equation}
for the classical action. Then the instanton contribution to the partition 
function becomes 
\begin{eqnarray}
\delta F
&\cong&
\frac{i}{4\sqrt{2\pi}\sqrt{NG(z_c)\sqrt{z_c-\alpha}
\left(\frac{3}{2}\sqrt{\mu}a\right)^{\frac{5}{2}}}}
\exp \left(-\frac{8\sqrt{3}\mu^{\frac{5}{4}}}{5g_s}\right)
\nonumber
\\
&=&
\frac{ig_s^{\frac{1}{2}}}
{8\cdot 3^{\frac{3}{4}}\cdot \sqrt{\pi}\mu^{\frac{5}{8}}}
\exp \left(-\frac{8\sqrt{3}\mu^{\frac{5}{4}}}{5g_s}\right).
\end{eqnarray}
Thus we have the chemical potential as a universal quantity which coincides 
with the one given in \cite{hanada}. 

\section{Conclusions and discussions}

In this paper we calculate the value of the chemical potential of the
D-instanton and obtain the universal value obtained in
\cite{hanada}. The problem was that the next to leading order contribution 
in eq.(\ref{Veff by M}) 
is divergent for $\beta \geq z\geq\alpha $. We discuss that this is
because for $z$ in such a region the existence of the eigenvalue affects
the distributions of other eigenvalues which cannot be treated
perturbatively. 

We should mention that although eq.(\ref{Veff by M}) is valid when
$z>\beta ,~\alpha >z$, we should take some care for $z$ close to 
$\alpha ,~\beta$. If one wants to calculate $V_{eff}(z)$ for some fixed
$z$ in the large $N$ limit, eq.(\ref{Veff by M}) can be used. However,
if one wants to know the value of $V_{eff}(z)$ for 
$z=\beta +{\cal O}(\frac{1}{N})$  for example, this equation may not be
used. As can be seen from eq.(\ref{inside}) the next to leading order
contribution to $V_{eff}(z)$ diverges in this case. Any way, we do not
have to be annoyed by these matters because we are interested in
$V_{eff}(z)$ around $z=\gamma$. 

Since we need only information from the Schwinger-Dyson equations to
calculate the chemical potential in our method, we will be able to
calculate it in other noncritical string theories. 
We would like to
study this problem elsewhere. 

\vspace{0.5cm}
\noindent
{\Large \bf Acknowledgments} \\
A. Y. acknowledges H. Fuji for valuable discussions.
This work is supported  in part by the Grants-in-Aid for Scientific Research 
13135224 and 13640308. 

\appendix
\section*{Appendix}

In this appendix, we will explain how to calculate $\int dze^{-V_{eff}}$ and 
get the result eq.(\ref{dzVeff}) in the large $N$ limit. 

In order to calculate $\int dze^{-V_{eff}(z)}$ to the next to leading order, 
we should be a little bit accurate. 
Let us fix the normalization of the integration 
measure $dM$ and $dM^\prime$ so that 
\begin{eqnarray}
\int dMe^{-\frac{N}{2g^2}{\rm Tr}\, M^2}
&=&
1,
\nonumber
\\
\int dM^\prime e^{-\frac{N-1}{2g^{\prime 2}}{\rm Tr}\, M^{\prime 2}}
&=&
1.
\label{normalization}
\end{eqnarray}
Then the matrix integrations are related to the integrations over the matrix 
eigenvalues as 
\begin{eqnarray}
\int dM
&=&
C_N^{-1}
\int\prod_{i=1}^{N}dz_i\prod_{i>j}(z_i-z_j)^2,
\nonumber
\\
\int dM^\prime 
&=&
C_{N-1}^{-1}
\int\prod_{i=1}^{N-1}dz^\prime_i\prod_{i>j}(z^\prime_i-z^\prime_j)^2,
\end{eqnarray}
where
\begin{eqnarray}
C_N
&=&
N! (2\pi)^{\frac{N}{2}}\left(\frac{g^2}{N}\right)^{\frac{N^2}{2}}
\prod_{k=1}^{N-1}k^{N-k},
\nonumber
\\
C_{N-1}
&=&
(N-1)! (2\pi)^{\frac{N-1}{2}}\left(\frac{g^{\prime 2}}{N-1}\right)^{\frac{(N-1)^2}{2}}
\prod_{k=1}^{N-2}k^{N-1-k}.
\end{eqnarray}
Now eq.(\ref{N/N-1}) becomes
\begin{equation}
\int dze^{-V_{eff}(z)}
=
\frac{C_N
\int dM
e^{-\frac{N}{g^2}{\rm Tr}\, V(M)}
}
{C_{N-1}
\int dM^\prime e^{-\frac{N-1}{g^{\prime 2}}{\rm Tr}\, V(M^\prime )}
}.
\end{equation}

With the normalization conditions eq.(\ref{normalization}), the partition 
functions can be expanded in the large $N$ limit as
\begin{eqnarray}
\int dM
e^{-\frac{N}{g^2}{\rm Tr}\, V(M)}
&=&
\exp \left(N^2F_0(g^2)+F_1(g^2)+\cdots \right),
\nonumber
\\
\int dM^\prime e^{-\frac{N-1}{g^{\prime 2}}{\rm Tr}\, V(M^\prime )}
&=&
\exp \left((N-1)^2F_0(g^{\prime 2})+F_1(g^{\prime 2})+\cdots \right),
\end{eqnarray}
where $F_n$ are contributions from the Feynman diagrams with the topology 
of genus $n$ Riemann surface. Therefore in the large $N$ limit, 
\begin{eqnarray}
\int dze^{-V_{eff}(z)}
&=&
\frac{C_N}{C_{N-1}}
\exp \biggl[(N-1)\left(2F_0(g^{\prime 2})
+g^{\prime 2}F_0^\prime (g^{\prime 2})\right)
\nonumber
\\
& &
\hspace{2cm}\left.
+F_0(g^{\prime 2})
+2g^{\prime 2}F_0^\prime (g^{\prime 2})
+\frac{1}{2}g^{\prime 4}F_0^{\prime\prime}(g^{\prime 2})
+{\cal O}\!\left(\frac{1}{N}\right)\right].
\nonumber
\\
& &
\label{answer}
\end{eqnarray}

$F_0(g^{\prime 2})$ can be evaluated using $\rho (z)$ in eq.(\ref{wzrho}). 
Since 
\begin{eqnarray}
\int dM^\prime e^{-\frac{N-1}{g^{\prime 2}}{\rm Tr}\, V(M^\prime )}
&=&
C_{N-1}^{-1}
\int\prod_{i=1}^{N-1}dz^\prime_i\prod_{i>j}(z^\prime_i-z^\prime_j)^2
e^{-\frac{N-1}{g^{\prime 2}}\sum V(z_i^\prime )}
\nonumber
\\
&=&
\exp 
\Biggl[(N-1)^2
\biggl(\int dz\int dz^\prime \rho (z)\rho(z^\prime )\ln |z-z^\prime |
\nonumber
\\
& &
\hspace{2.5cm}
-\frac{1}{g^{\prime 2}}\int dz\rho (z)V(z)
+\frac{3}{4}-\frac{1}{2}\ln g^{\prime 2}\biggr)
\nonumber
\\
& &
\hspace{3cm}
+{\cal O}(N^0)\Biggr],
\end{eqnarray}
we get
\begin{equation}
F_0(g^{\prime 2})
=
\int dz\int dz^\prime \rho (z)\rho(z^\prime )\ln |z-z^\prime |
-\frac{1}{g^{\prime 2}}\int dz\rho (z)V(z)
+\frac{3}{4}-\frac{1}{2}\ln g^{\prime 2}.
\label{F_0}
\end{equation}
From eq.(\ref{wzrho}), we obtain 
\begin{equation}
\int dz^\prime \rho (z^\prime )\ln |z-z^\prime |
=
\lim_{\Lambda\rightarrow\infty}
\left[\int_\Lambda^z dz^\prime \mbox{Re}~ w(z^\prime ) +\ln \Lambda \right].
\end{equation}
Substituting this formula into eq.(\ref{F_0}), taking the form of $w(z)$ in 
eq.(\ref{wz}) into account, we get 
\begin{equation}
F_0(g^{\prime 2})
=
-\frac{1}{2g^{\prime 2}}\int dz\rho (z)V(z)
+R+\frac{3}{4}-\frac{1}{2}\ln g^{\prime 2},
\label{F_0R}
\end{equation}
where $R$ is defined in eq.(\ref{R}). 

In order to obtain $\int dze^{-V_{eff}}$, we should calculate 
$2F_0(g^{\prime 2})+g^{\prime 2}F_0^\prime (g^{\prime 2})$ and 
$F_0(g^{\prime 2})
+2g^{\prime 2}F_0^\prime (g^{\prime 2})
+\frac{1}{2}g^{\prime 4}F_0^{\prime\prime}(g^{\prime 2})$. 
After substituting eq.(\ref{F_0R}) into these, it is convenient to proceed as 
follows. From the relation
\begin{eqnarray}
&&g^{\prime 2}\partial_{g^{\prime 2}}
\ln \left[
\int\prod_{i=1}^{N-1}dz^\prime_i\prod_{i>j}(z^\prime_i-z^\prime_j)^2
e^{-\frac{N-1}{g^{\prime 2}}\sum V(z_i^\prime )}
\right]\nonumber\\
&&\hspace{40pt}=
\frac{(N-1)^2}{g^{\prime 2}}\int dz\rho (z)V(z)+{\cal O}((N-1)^0),
\end{eqnarray}
one can prove the following identity:
\begin{equation}
g^{\prime 2}\partial_{g^{\prime 2}}
\left(R-\frac{1}{2g^{\prime 2}}\int dz\rho (z)V(z)\right)
=
\frac{1}{g^{\prime 2}}\int dz\rho (z)V(z).
\end{equation}
Using this identity, we can express the two quantities above in terms of 
$R$ as
\begin{eqnarray}
2F_0(g^{\prime 2})+g^{\prime 2}F_0^\prime (g^{\prime 2})
&=&
2R+1-\ln g^{\prime 2},
\label{leading}
\\
F_0(g^{\prime 2})
+2g^{\prime 2}F_0^\prime (g^{\prime 2})
+\frac{1}{2}g^{\prime 4}F_0^{\prime\prime}(g^{\prime 2})
&=&
R+g^{\prime 2}\partial_{g^{\prime 2}}R-\frac{1}{2}\ln g^{\prime 2}.
\label{subleading}
\end{eqnarray}
From eq.(\ref{leading}), one can get the leading order part of 
eq.(\ref{dzVeff}). 

To further simplify the right hand side of eq.(\ref{subleading}), 
we use the following expression for $R$:
\begin{equation}
R
=
\frac{1}{N-1}\left\langle {\rm Tr}\ln (\beta -M^\prime )\right\rangle 
-\frac{1}{2g^{\prime 2}}V(\beta ).
\end{equation}
From this expression, we get 
\begin{eqnarray}
g^{\prime 2}\partial_{g^{\prime 2}}R
&=&
\frac{1}{2g^{\prime 2}}V(\beta )
+
\left\langle \frac{1}{g^{\prime 2}}{\rm Tr}\, V(M^\prime ){\rm Tr}
\ln (\beta -M^\prime )\right\rangle_c 
\nonumber
\\
& &
\hspace{5mm}
+
\left(\frac{1}{N-1}\left\langle {\rm Tr}\frac{1}
{\beta -M^\prime }\right\rangle 
-\frac{1}{2g^{\prime 2}}V^\prime (\beta )\right)
g^{\prime 2}\partial_{g^{\prime 2}}\beta 
\nonumber
\\
&=&
\frac{1}{2g^{\prime 2}}V(\beta )
+
\left\langle \frac{1}{g^{\prime 2}}{\rm Tr}\, V(M^\prime ){\rm Tr}
\ln (\beta -M^\prime )\right\rangle_c ,
\end{eqnarray}
where we used eq.(\ref{wz}) in the last line. 
Therefore we obtain the following expression for 
$R+g^{\prime 2}\partial_{g^{\prime 2}}R$:
\begin{eqnarray}
&&\hspace{-40pt}R+g^{\prime 2}\partial_{g^{\prime 2}}R\nonumber\\
&=&
\frac{1}{N-1}\left\langle {\rm Tr}\ln(\beta -M^\prime )\right\rangle 
+
\left\langle \frac{1}{g^{\prime 2}}{\rm Tr}\, V(M^\prime ){\rm Tr}
\ln (\beta -M^\prime )\right\rangle_c 
\nonumber
\\
&=&
\lim_{\Lambda\rightarrow\infty}
\left[\int_\Lambda^\beta dz
\left(w(z)+
\oint_C \frac{dz^\prime}{2\pi i}
\frac{1}{g^{\prime 2}}V(z^\prime )
\left\langle {\rm Tr}\frac{1}{z-M^\prime}{\rm Tr}
\frac{1}{z^\prime -M^\prime}\right\rangle_c \right)+\ln\Lambda
\right].\nonumber\\
&&
\label{R+dR}
\end{eqnarray}
Here we have chosen $C$ to be a closed contour in the $z$ plane, which 
encloses the cut $[\alpha ,\beta ]$ leaving outside the point $z^\prime$. 

Now let us substitute the correlation functions $w(z)$ and 
$\langle {\rm Tr}\frac{1}{z-M^\prime}{\rm Tr}\frac{1}{z^\prime -M^\prime}\rangle $ of the 
matrix model into this formula. It is known that 
$w(z)$ has the following 
expression\cite{Migdal:1984gj} \cite{makeenko}: 
\begin{eqnarray}
w(z)
&=&
\oint_C \frac{dz^\prime}{4\pi i}
\frac{1}{g^{\prime 2}}
\frac{V^\prime (z^\prime )}{z-z^\prime }
\sqrt{\frac{(z-\alpha )(z-\beta )}{(z^\prime -\alpha )(z^\prime -\beta
)}}.\nonumber\\
& &
\end{eqnarray}
On the other hand, 
$\left\langle {\rm Tr}\frac{1}{z-M^\prime}{\rm Tr}
\frac{1}{z^\prime -M^\prime}\right\rangle_c $ 
in eq.(\ref{two loop}) can be rewritten as
\begin{equation}
\left\langle {\rm Tr}\frac{1}{z-M^\prime}{\rm Tr}
\frac{1}{z^\prime -M^\prime}\right\rangle_c  
=
-\frac{1}{2}\partial_{z^\prime}
\left[\frac{1}{z-z^\prime}
\left(1-\sqrt{\frac{(z^\prime -\alpha )(z^\prime -\beta )}{(z-\alpha )(z-\beta )}}\right)
\right].
\end{equation}
Moreover, it is known that the following identities 
hold\cite{Migdal:1984gj}\cite{makeenko}:
\begin{eqnarray}
\oint_C \frac{dz^\prime}{2\pi i}
\frac{1}{g^{\prime 2}}
\frac{V^\prime (z^\prime )}{\sqrt{(z^\prime -\alpha )(z^\prime -\beta )}}
&=&
0,
\label{id1}
\\
\oint_C \frac{dz^\prime}{2\pi i}
\frac{1}{g^{\prime 2}}
\frac{z^\prime V^\prime (z^\prime )}
{\sqrt{(z^\prime -\alpha )(z^\prime -\beta )}}
&=&
2.
\label{id2}
\end{eqnarray}

Substituting the expressions for the correlators above into eq.(\ref{R+dR}), 
taking the identities eqs.(\ref{id1})(\ref{id2}) into 
account, we eventually get
\begin{eqnarray}
R+g^{\prime 2}\partial_{g^{\prime 2}}R
&=&
\lim_{\Lambda\rightarrow\infty}
\left[\int_\Lambda^\beta dz \frac{1}{\sqrt{(z-\alpha )(z-\beta )}}+\ln\Lambda \right]
\nonumber
\\
&=&
\ln \frac{\beta -\alpha }{4}. 
\end{eqnarray}

Substituting all these results into eq.(\ref{answer}), we get \begin{equation}
\int dze^{-V_{eff}}
\cong 
\frac{\pi (N-1)}{2}(\beta -\alpha )e^{2(N-1)R}.
\end{equation}

\end{document}